\documentstyle[pra,aps,multicol,psfig]{revtex}
\draft
\begin{document}
\title{Collisionless modes of a trapped Bose gas}
\author{M.J. Bijlsma and H.T.C. Stoof \\
Institute for Theoretical Physics, University of Utrecht, \\
Princetonplein 5, 3584 CC Utrecht, The Netherlands}
\maketitle
\begin{abstract}
We calculate the excitation frequencies of the 
low-lying modes of a trapped Bose-condensed gas
at nonzero temperatures. 
We include in our calculation the dynamics of the noncondensed cloud,
and find agreement with experimental results
if we assume that in the experiment both the in-phase and 
out-of-phase monopole modes are excited simultaneously.
In order to explore whether this is indeed the correct explanation, 
we also calculate how strongly the modes couple to a 
perturbation of the external trapping potential.
\end{abstract}
\begin{multicols}{2}
\section{INTRODUCTION}
Since the experimental realization of Bose-Einstein condensation
in magnetically trapped $^{87}$Rb, $^{7}$Li and $^{23}$Na
\cite{anderson,bradley,davis} gases, various properties
of these Bose-condensed systems have been studied
both experimentally and theoretically.
A challenging problem that has attracted much attention,
but is still only partially resolved,
is the temperature dependence 
of the excitation frequencies of the collective modes 
of a harmonically confined Bose-condensed gas.
In principle, it's solution requires 
solving the coupled dynamics of  
the condensed and noncondensed parts of a highly
inhomogeneous interacting gas \cite{proukakis2}. 

In general, the 
dynamics of an interacting many-body system  
can be classified as being either in the 
hydrodynamic or in the collisionless limit,
and the nature of the collective excitations 
of the system changes as one goes from
one limit to the other \cite{kadanoff}.
In the hydrodynamic regime
the mean free path of the 
quasiparticles is small relative to the wavelength of the collective
excitations. Therefore one can assume the system to be in local
equilibrium. In the case of a Bose-condensed gas,
this important assumption leads 
to a description of the coupled dynamics of the condensate 
and the noncondensate clouds by the well-known Landau two-fluid equations.
The solution to these equations have been studied by several authors
\cite{zaremba,shenoy,kavoulakis1},
and also experimentally one is trying to probe this regime \cite{stamper-kurn}.
In contrast, in the collisionless regime there is no local equilibrium
because the mean free path of the  
quasiparticles is much larger than the wavelength of the collective
modes. Up to now, most experiments with Bose-condensed gases
are performed in this
regime. What is of prime interest at present 
is that these experiments, performed in an
axially symmetric trapping potential, indicate 
a strong temperature dependence of the excitation frequencies for
temperatures relatively close to the critical temperature $T_{c}$ \cite{jin2}.
The first attempts to explain this temperature dependence
theoretically were unsatisfactory for a number of reasons.

Far below the critical temperature,
the collisionless modes of a trapped Bose gas are
accurately described by the Gross-Pitaevskii equation.
This equation is a nonlinear Schr\"odinger equation
that describes the time evolution of the condensate wave function 
at temperatures where the 
noncondensate density is negligible. It is nonlinear because it
includes the mean-field interaction of the condensate with itself.
Theoretical calculations solving this equation 
\cite{stringari,singh,edwards,castin,perez-garcia,dodd96,javanainen,you,ohberg}, 
are in good agreement with measurements of the low-lying 
excitation frequencies in this temperature regime \cite{jin1,mewes}. 

However, at higher temperatures the noncondensate fraction
becomes large. Therefore,
one has to include into the nonlinear Schr\"odinger equation
also the effect of the mean-field interaction 
of the thermal cloud with the condensate.
The low-lying excitation frequencies predicted
by the resulting nonlinear Schr\"odinger equation 
have been found numerically and 
show almost no temperature dependence \cite{hutchinson1,dodd}.
The lack of temperature dependence can partly be cured
by including in the effective two-particle interaction the 
many-body effects of the surrounding gas on the collisions.
This causes the interaction to 
become strongly temperature dependent \cite{bijlsma} and 
the frequencies of the low-lying modes then also 
depend on temperature \cite{hutchinson2}. The frequency of 
the mode with azimuthal quantum number $m=2$ 
found in this way agrees with experiments quite well.
This is, however, not the case for the $m=0$ mode. 

The latter is related to a fundamental problem with the nonlinear
Schr\"odinger equation used, which 
describes the dynamics of the condensate in the
presence of a static noncondensed cloud. As a result 
it violates the generalized Kohn theorem, 
which states that for a harmonically confined
many-particle system there are exactly three
center-of-mass modes with excitation frequencies equal to
the three trapping frequencies.
Clearly, this is caused by the fact that
we also have to describe the time evolution 
of the thermal cloud.
Therefore, we proposed to describe 
the collective excitations in the collisionless regime
by a nonlinear Schr\"odinger equation 
for the condensate wave function, which is coupled 
to a collisionless Boltzmann or Vlasov-Landau equation,
describing the dynamics of the noncondensed cloud \cite{stoof,kirkpatrick}. 
The resulting theory contains the Kohn modes exactly.

Even above the critical temperature,
solving the collisionless Boltzmann equation
for the collective modes of the gas is not straightforward.
As an illustration, we note that this equation also
contains the hydrodynamic modes as a special 
class of solutions. However, we are in particular interested
in temperatures relatively close to the critical temperature, 
where the discrepancy between theory and experiment
is largest. Therefore, we want to formulate a manageable 
theory that has the correct behavior both near zero temperature and 
near the critical temperature. With this objective in mind,
we can treat 
the thermal cloud in the Hartree-Fock approximation, 
because near the critical temperature 
the mean-field interaction is small
compared to the average kinetic energy of the noncondensed atoms. 
Moreover, near zero temperature the presence of the
thermal cloud is unimportant.

Finally, to solve the resulting set of coupled partial 
differential equations that describe the dynamics of the
trapped Bose-condensed gas in the time-dependent 
Hartree-Fock approximation, we 
apply a dynamical scaling of the ideal gas results
for both the condensate
wave function and the Wigner distribution describing
the noncondensed cloud. 
  
The paper is organized as follows. In Sec. 
\ref{section2} we present the general theoretical 
framework that describes collisionless dynamics of
a trapped Bose-condensed gas. In Sec. \ref{section3}
we show how the excitation frequencies of the low-lying 
modes can be determined by means of a dynamical scaling {\it ansatz}.
We give a physical motivation for our {\it ansatz}
and argue in particular that it is accurate near the critical temperature.
In Sec. \ref{section4} we show how the linear response
to an external perturbation of the trapping frequency is calculated 
in our approach. This calculation is performed for
reasons that become clear 
later on in Sec. \ref{section5}, when
we compare our results with the experiment.
We end the paper in Sec. \ref{section6}
with a summary of our conclusions and an outlook. 
\section{COLLISIONLESS DYNAMICS}
\label{section2}
In the Heisenberg picture, the quantum-mechanical evolution 
of a many-body system of Bose particles in an external trapping potential  
can be described by a field 
operator $\hat \psi({\bf x},t)$ and a Hamiltonian

\begin{eqnarray}
\hat H & = & \left. \int \! d{\bf x} \right[ \hat \psi^{\dagger}({\bf x},t) 
{\cal H}_{0} \hat \psi({\bf x},t) + \nonumber \\ 
& & \hspace{-0.35cm} \frac{1}{2} \left. \int \! d{\bf x'} \; 
\hat \psi^{\dagger}({\bf x},t) 
\hat \psi^{\dagger}({\bf x'},t) 
V\!({\bf x\!-\!x'}) 
\hat \psi({\bf x'},t) 
\hat \psi({\bf x},t) \right] \; ,
\end{eqnarray}   
where $V({\bf x-x'})$ is the two particle interaction,
the single particle Hamiltonian ${\cal H}_{0}$ is given by

\begin{eqnarray}
{\cal H}_{0} & = & -\frac{\hbar^{2} \nabla^{2}}{2 m} + V_{ext}({\bf x}) 
 \; ,
\end{eqnarray}
and $V_{ext}({\bf x})=\sum_{i} m \omega_{i}^{2} x_{i}^{2}/2$ denotes 
the external trapping potential.
The field operator satisfies the equal-time commutation relations

\begin{eqnarray}
\left[ \hat \psi({\bf x},t), \hat \psi({\bf x'},t) \right] & = & 0 \nonumber 
\; , \\
\left[ \hat \psi({\bf x},t), \hat \psi^{\dagger}({\bf x'},t) \right] & = & 
\delta({\bf x-x'}) \; .
\end{eqnarray}

By definition, the time evolution of the field operators is
determined by the Heisenberg equation of motion

\begin{eqnarray}
i \hbar \frac{\partial \hat \psi({\bf x},t)}{\partial t} & = &
\left[ \hat \psi({\bf x},t), \hat H \right] \; .
\end{eqnarray}
From this equation 
we can now easily derive an equation of motion for 
the condensate wave function $\Psi({\bf x},t)$, which is 
defined as the
expectation value $\langle \hat \psi({\bf x},t) \rangle$.
Applying a mean-field approximation 
results in a nonlinear Schr\"odinger equation
for $\Psi({\bf x},t)$,
which includes the mean-field 
interaction with the noncondensate density $n'({\bf x},t)$.
It reads

\begin{eqnarray}
\label{nlse}
i \hbar \frac{\partial \Psi({\bf x},t)}{\partial t} & = &
\left\{
-\frac{\hbar^{2} \nabla^{2}}{2m} + V_{ext}({\bf x}) \right. 
\nonumber \\
& & \left. \mbox{} \; \; + 
T^{2B} \left[ 2 n'({\bf x},t) + n_{0}({\bf x},t) \right]
\frac{}{\mbox{}} \!\!  \right\} \Psi({\bf x},t) \; .
\end{eqnarray}
Here, the condensate density $n_{0}({\bf x},t)$ is defined as the
modulus squared of the condensate wave function,
i.e., $|\Psi({\bf x},t)|^{2}$. 
The factor of two difference between the condensate and noncondensate
mean-field interactions is understood physically by realizing
that the noncondensate contributes both a Hartree and a Fock term,
but the condensate only a Hartree term. 
Furthermore, the two-body interaction has been 
renormalized to a hard-core potential 
$T^{2B} \delta({\bf x}- {\bf x}')$,
where the two-body scattering matrix $T^{2B}=4 \pi \hbar^{2} a/m$ solves
the Lipmann-Schwinger equation for the scattering of two 
particles with zero momentum and $a$ denotes the interatomic 
scattering length. In principle, the anomalous average
has renormalized the two-body interaction potential to the many-body
T-matrix $T^{MB}$, which also includes the effect of the surrounding
gas on the collisions between two particles \cite{proukakis}.
For simplicity, however, we will use the two-body scattering matrix
instead, neglecting the effective 
temperature dependence of the two-body interactions.

Next, we want to derive an equation of motion for
the noncondensed cloud. This is achieved by
considering the Heisenberg equation of motion for the one-particle
density matrix,
\begin{eqnarray}
\label{one_particle_density_matrix} 
\left(
\begin{array}{cc}
\langle \hat \psi'^{\dagger}({\bf x},t) \hat \psi'({\bf x'},t) \rangle & 
\langle \hat \psi'({\bf x},t) \hat \psi'({\bf x'},t) \rangle \\ 
\langle \hat \psi'^{\dagger}({\bf x},t) \hat \psi'^{\dagger}({\bf x'},t) 
\rangle & \langle \hat \psi'({\bf x},t) \hat 
\psi'^{\dagger}({\bf x'},t) \rangle 
\end{array}
\right)
\; .
\end{eqnarray}
Here, the field operator describing the 
noncondensate is defined according to  
$\hat \psi'({\bf x},t) \equiv \hat \psi({\bf x},t) - \Psi({\bf x},t)$.

From Eq.~(\ref{one_particle_density_matrix}) we
derive a Boltzmann equation.
This Boltzmann equation describes the time evolution
of the quasiparticle distribution function $F({\bf k},{\bf x},t)$,
which is related to the Fourier transform 
of the above one-particle density matrix by a
local Bogoliubov transformation.
More precisely, it is derived
from the equations of motion for the one-particle 
density matrix by performing a gradient expansion
in the center-of-mass coordinate $({\bf x+x'})/2$. 
This is justified because in general the 
noncondensate density profile varies on
a much larger length scale 
than the external trapping potential. 
The resulting Boltzmann equation contains two steaming terms,
corresponding to the local group velocity of a quasiparticle  
and the local force on a quasiparticle. The velocity and the force
are given by the momentum and the spatial derivative of 
the energy dispersion $E({\bf k},{\bf x},t)$, respectively. 
The distribution of quasiparticles can also change 
because of collisions and on the right-hand side there
is an associated collision term. In total the Boltzmann equation
therefore reads

\begin{eqnarray}
\label{cbe}
\left[
\frac{\partial}{\partial t} + 
\frac{\partial E}{\partial \hbar {\bf k}} 
\cdot \frac{\partial}{\partial {\bf x}} -  
\frac{\partial E}{\partial {\bf x}}  
\cdot \frac{\partial}{\partial \hbar {\bf k}}  
\right] F & = & 
\left[ 
\frac{\partial F}{\partial t} 
\right]_{coll.} \; .
\end{eqnarray}
If we treat the quasiparticles in the Popov approximation
\cite{popov},
their energy in the frame where the superfluid velocity $v_{s}({\bf x},t)$
is zero, is given by
\begin{eqnarray}
\label{popov}
E({\bf x},{\bf k},t) & = & \left(
\left[  \frac{\hbar^{2} {\bf k}^{2}}{2m} + V_{ext}({\bf x})
+ 2T^{2B}n({\bf x},t) \right. \right. \nonumber \\ 
& & \left. \left. \hspace{1.2cm} \mbox{} + \hbar \dot \theta({\bf x},t) 
+ \frac{m v_{s}^{2}({\bf x},t)}{2}
\right]^{2} \right. \nonumber \\
& & \left. \mbox{} - \left[ T^{2B} n_{0}({\bf x},t)  
\right]^{2} \right)^{1/2} 
+ \hbar {\bf k} \cdot {\bf v}_{s}({\bf x},t)
\; .
\end{eqnarray}
Here, $\theta({\bf x},t)$ is the phase of the condensate wave function,
i.e., $\Psi({\bf x},t)=\sqrt{n_{0}({\bf x},t)} \exp{i \theta({\bf x},t)}$.
Furthermore, 
the superfluid velocity is proportional to the gradient of that phase
or more precisely 
${\bf v}_{s}({\bf x},t)=\hbar \nabla \theta({\bf x},t)/m$. 

In the temperature region of interest it is sufficient
to treat the quasiparticles in the Hartree-Fock approximation.
The energy in the laboratory frame is then given by

\begin{eqnarray}
\label{hf}
E({\bf x},{\bf k},t)=\frac{\hbar^{2} {\bf k}^{2}}{2m} 
+ V_{ext}({\bf x}) + 2 T^{2B} n({\bf x},t)  
\; .
\end{eqnarray}
The resulting nonlinear Schr\"odinger  and collisionless Boltzmann
equation are expected to describe the dynamics of the trapped Bose gas
near the critical temperature $T_{c}$,
because in this temperature region the mean-field interaction of the
condensate is small compared to the 
average kinetic energy of the noncondensed cloud,
implying that the Hartree-Fock approximation is valid. 
Moreover, they will also be accurate near zero temperature, because
here the noncondensate density is negligible.
\section{VARIATIONAL APPROACH}
\label{section3}
Directly solving the set of
equations that describe the coupled collisionless dynamics 
of the condensate and the noncondensate,
i.e., the nonlinear Schr\"odinger equation given by Eq. (\ref{nlse}) and 
the collisionless Boltzmann equation following from 
Eq. (\ref{cbe}), is difficult
even when treating the quasiparticles in the Hartree-Fock approximation.
However, it is well known that a simple scaling {\it ansatz} for
the condensate wave function gives   
the correct frequencies of the low-lying modes at zero temperature
\cite{castin,perez-garcia}.
Therefore, we use a similar method to solve our set of equations
and also assume the time dependence of 
the quasiparticle distribution function 
to be given by a dynamical scaling {\it ansatz}.
We thus get 

\begin{mathletters}
\begin{eqnarray}
n_{0}({\bf x},t) & = & \left[ \prod_{j}\frac{1}{\lambda_{j}} \right]
\Phi_{n_{0}} \left( 
\left\{ 
\frac{x_{i}-\eta_{i}^{(1)}}{\lambda_{i}} 
\right\}
\right) \; , 
\end{eqnarray}
\begin{eqnarray}
\theta({\bf x},t) & = & \sum_{i} \frac{m \omega_{i}}{\hbar} 
\left[ \eta_{i}^{(3)} x_{i} + \beta_{i} x_{i}^{2} \right] \; , 
\end{eqnarray}
\begin{eqnarray}
F({\bf k},{\bf x},t) & = & \! \left[ \! \prod_{j} \! c_{j} \! \right] 
\! \Phi_{F} \! \left( \! \!
\left\{
\sqrt{c_{i}} \left[ \frac{x_{i}-\eta_{i}^{(2)}}{\alpha_{i}} \right]
\right\},
\right. \\
& & \left. 
\left\{
\sqrt{c_{i}} \alpha_{i} 
\! \left[ 
k_{i} \!-\! \frac{m}{\hbar} \frac{\dot \alpha_{i}}{\alpha_{i}}
[x_{i} \!-\! \eta_{i}^{(2)}] \!-\!  
\frac{m}{\hbar} \dot \eta_{i}^{(2)} \! \right] \! 
\right\}
\right) \; .
\nonumber 
\end{eqnarray}
\end{mathletters}

The six parameters $\{ \lambda_{i} \}$ and $\{ \alpha_{i} \}$ 
describe the coupled monopole and quadrupole oscillations
of the two components of the gas.
The six parameters $\{ \eta_{i}^{(1)} \}$ and $\{ \eta_{i}^{(2)} \}$
are included to describe the in-phase and out-of-phase dipole, or Kohn modes. 
Notice that the momentum argument of the scaling {\it ansatz} 
for the distribution function contains 
explicit time derivatives of the scaling parameters.
Together with the quadratic form of the phase,
this ensures that the continuity
equation is to be satisfied.
The variational parameters $\{\eta_{i}^{(3)}\}$ and
$\{\beta_{i}\}$ are therefore related to
$\{\eta_{i}^{(1)}\}$ and $\{\lambda_{i}\}$ 
in a way that is specified below.
Due to the mean-field interaction, the equilibrium
values of $\{ \lambda_{i} \}$
and $\{ \alpha_{i} \}$ are in general not equal to one. 
In order to ensure that the equilibrium
profile is isotropic in momentum space,
the factors $\{ c_{i} \}$ are inserted,
which are equal to $1/\bar \alpha_{i}^{2}$,
where $\bar \alpha_{i}$ denotes the equilibrium 
value of $\alpha_{i}$.
The modes of the noncondensed cloud described
by this {\it ansatz}, are however not constrained to be
isotropic in momentum space. 
Hence they describe more complicated dynamics 
than hydrodynamic motion, as desired in the collisionless limit.

The equations of motion for the scaling parameters of the 
condensate are  
most easily found from the lagrangian density for the nonlinear Schr\"odinger
equation written in terms of the density and the phase,

\begin{eqnarray}
\label{lagrangian}
{\cal L} & = & \hbar n_{0}({\bf x,t}) 
\frac{\partial \theta({\bf x,t})}{\partial t} +
\frac{\hbar^{2}}{2m} 
\frac{[\nabla n_{0}({\bf x,t})]^{2}}{4 n_{0}({\bf x,t})} +
\nonumber \\
& & \frac{\hbar^{2}}{2m} 
n_{0}({\bf x,t}) 
[\nabla \theta({\bf x,t})]^{2} +
\\
& & \left[
V_{ext}({\bf x}) + 2 T^{2B} n'({\bf x},t)
+ \frac{T^{2B}}{2} n_{0}({\bf x,t}) \right] n_{0}({\bf x,t}) 
\nonumber \; .
\end{eqnarray}
Inserting the scaling {\it ansatz} for the density and 
the appropriate phase into this lagrangian density,
and integrating over space yields a lagrangian for the
scaling parameters. The equations of motion for these parameters are
then given by the appropriate Euler-Lagrange equations.
The equations of motion for the scaling parameters of the noncondensate
are found by taking moments of the collisionless Boltzmann equation
with respect to 
$k_{i}, x_{i}, k_{i}k_{j}, x_{i}k_{i}$ and $x_{i}x_{j}$.  

The resulting equations of motion for the 
in total $12$ variational parameters are
\begin{mathletters}
\label{evolution}
\begin{eqnarray}
\ddot \eta_{i}^{(1)} + \omega_{i}^{2} \eta_{i}^{(1)} & = & 
- \frac{2 T^{2B}}{m N_{0}} 
\frac{\partial}{\partial \eta_{i}^{(1)}}
\langle \Phi_{n'}(\mbox{\boldmath $\xi$}_{1}) \rangle_{c} 
\prod_{j} \frac{\sqrt{c_{j}}}{\alpha_{j}} \; ,
\end{eqnarray}
\begin{eqnarray}
\ddot \eta_{i}^{(2)} + \omega_{i}^{2} \eta_{i}^{(2)} & = &  
- \frac{2 T^{2B}}{m N'} 
\frac{\partial}{\partial \eta_{i}^{(2)}}
\langle \Phi_{n_{0}}(\mbox{\boldmath $\xi$}_{2}) \rangle_{nc}
\prod_{j} \frac{1}{\lambda_{j}} \; , 
\end{eqnarray}
\begin{eqnarray}
\ddot \lambda_{i} + \omega_{i}^{2} \lambda_{i} & = & 
\frac{T_{i}^{kin,c}}
{m \langle x_{i}^{2} \rangle_{c}} \frac{2}{\lambda_{i}^{3}} +
\frac{T^{2B}
\langle \Phi_{n_{0}}({\bf x}) \rangle_{c}}
{2 m \langle x_{i}^{2} \rangle_{c}}
\frac{1}{\lambda_{i}} 
\prod_{j} \frac{1}{\lambda_{j}} 
\nonumber \\
&  & 
- \frac{2 T^{2B}}{m \langle x_{i}^{2} \rangle_{c}} 
\frac{\partial}{\partial \lambda_{i}}
\langle \Phi_{n'}(\mbox{\boldmath $\xi$}_{1}) 
\rangle_{c} \prod_{j} \frac{\sqrt{c_{j}}}{\alpha_{j}} 
\; , 
\end{eqnarray}
\begin{eqnarray}
\ddot \alpha_{i} + \omega_{i}^{2} \alpha_{i} & = &
\frac{T_{i}^{kin,nc}}{m \langle x_{i}^{2} \rangle_{nc}}
\frac{2}{\alpha_{i}^{3}} + 
\frac{T^{2B} \langle \Phi_{n'}({\bf x}) \rangle_{nc}}
{m \langle x_{i}^{2} \rangle_{nc}}
\frac{c_{i}}{\alpha_{i}} 
\prod_{j} \frac{\sqrt{c_{j}}}{\alpha_{j}} 
\nonumber \\
&  & 
- \frac{2 T^{2B}}{m \langle x_{i}^{2} \rangle_{nc}}
\frac{\partial}{\partial \alpha_{i}}
\langle \Phi_{n_{0}}(\mbox{\boldmath $\xi$}_{2}) \rangle_{nc}
\prod_{j} \frac{1}{\lambda_{j}} \; ,
\end{eqnarray}
\end{mathletters}
whereas the dynamics of the phase is determined by

\begin{mathletters}
\begin{eqnarray}
\beta_{i} & = & \frac{1}{2 \omega_{i}} 
\frac{\dot \lambda_{i}}{\lambda_{i}} \; , 
\end{eqnarray}
\begin{eqnarray}
\eta_{i}^{(3)} & = &  -2 \beta_{i} \eta_{i}^{(1)} + 
\frac{\dot \eta_{i}^{(1)}}{\omega_{i}} \; . 
\end{eqnarray}
\end{mathletters}
For notational convenience we have defined
the arguments $\mbox{\boldmath $\xi$}_{1}$ and 
$\mbox{\boldmath $\xi$}_{2}$ by

\begin{mathletters}
\begin{eqnarray}
\xi_{1,i} & = & (\lambda_{i} x_{i} + \eta_{i}^{(1)} - 
\eta_{i}^{(2)}) \sqrt{c_{i}}/ \alpha_{i} \; ,
\end{eqnarray}
\begin{eqnarray} 
\xi_{2,i} & = & (\alpha_{i} x_{i}/ \sqrt{c_{i}} + \eta_{i}^{(2)} - 
\eta_{i}^{(1)})/ \lambda_{i} \; . 
\end{eqnarray}
\end{mathletters}
Also, the kinetic energy 
of the condensate and noncondensate are given by

\begin{eqnarray}
T_{i}^{kin,c} & = & \int \! d{\bf x} \; \frac{\hbar^{2}}{2m} 
\frac{[\partial_{i} \Phi_{n_{0}}({\bf x})]^{2}}
{4 \Phi_{n_{0}}({\bf x})} \; , 
\end{eqnarray}
and 
\begin{eqnarray}
T_{i}^{kin,nc} & = & \int \! d{\bf x} \int \! {d{\bf k} \over (2 \pi)^{3}} \;
\frac{\hbar^{2} k_{i}^{2}}{2m} \Phi_{F}({\bf x},{\bf k}) \; ,
\end{eqnarray}
respectively. 
Finally, the weighted averages with respect to the equilibrium condensate 
and noncondensate densities are  

\begin{eqnarray}
\langle f({\bf x}) \rangle_{c} & = & \int \! d{\bf x} \; 
\Phi_{n_{0}}({\bf x}) f({\bf x}) \; , 
\end{eqnarray}
and
\begin{eqnarray}
\langle f({\bf x}) \rangle_{nc} & = & 
\int \! d{\bf x} \;  
\Phi_{n'}({\bf x}) f({\bf x}) \; ,
\end{eqnarray}
where $\Phi_{n'}({\bf x})\!=\!(2\pi)^{-3} \! 
\int d{\bf k} \; \Phi_{F}({\bf k, x})$.
Note that the equations of motion for $\{ \lambda_{i} \}$
and $\{ \alpha_{i} \}$ are of the same form, apart from
a factor of two difference between the mean-field interaction 
of the condensate with itself and the mean-field interaction of 
the noncondensate with itself.
Of course, this is the same factor of two difference
that was mentioned above in relation to the nonlinear 
Schr\"odinger equation.

In the Thomas-Fermi limit
we can neglect the average kinetic energy 
of the condensate relative to the mean-field interactions,
and the ground-state density profile of the condensate
is approximately an inverted parabola. The opposite limit, where 
the mean-field interaction is much less than the average
kinetic energy, results in 
a gaussian density profile. 
For the experiments of interest here, a gaussian profile is
appropriate near the critical temperature.
At low temperatures this is no longer true
and an inverted parabola is more accurate.
However, for our purposes it is convenient
to  take at all temperatures a gaussian {\it ansatz} 
for the condensate, because
it is known to give the correct frequencies even at zero temperature
\cite{perez-garcia}. Hence,  

\begin{eqnarray}
\label{explicit1}
\Phi_{n_{0}}({\bf x}) & = & 
N_{0} \prod_{j} \left( 
\frac{m \omega_{j}}{\hbar \pi}
\right)^{1/2} 
e^{
- \sum_{i} \frac{m \omega_{i}}{\hbar} x_{i}^{2}} \; .
\end{eqnarray}

Since the effect of the noncondensate is
most important near $T_{c}$,
we take the dynamical scaling {\it ansatz}
for the quasiparticle distribution 
to be a Bose function. 
Near the critical temperature
and neglecting mean-field interactions, 
this indeed gives the correct mode frequencies, i.e., $\{ 2 \omega_{i} \}$. 
We thus take

\begin{eqnarray}
\label{explicit2}
\Phi_{F}({\bf k},{\bf x}) & = & 
\frac{N' \prod_{j} \! \left( \beta \hbar \omega_{j} \right)}{\zeta(3)}
N \! \left( \frac{\hbar^{2} k^{2}}{2 m} \! + \!
\sum _{i} \! \frac{m \omega_{i}^{2} x_{i}^{2}}{2} \! \right) \; ,
\end{eqnarray}  
where $N(\varepsilon)$ is the usual Bose-distribution function

\begin{eqnarray}
N(\varepsilon) = \left[ e^{\beta \varepsilon} - 1 \right]^{-1} \; .
\end{eqnarray}
Here, $\beta=1/k_{B} T$ and $\zeta(3) \approx 1.202$.
The normalization factors are such that the total number of
condensed and noncondensed atoms is equal to $N_{0}$ and $N'$, respectively.
We take the temperature dependence of the total number of condensed atoms
to be given by the noninteracting result, $N_{0}=[1-(T/T_{c})^{3}] N$,
where $N = N_{0}+N'$ is the total number of particles.

The equations of motion for the scaling parameters
that result from inserting Eqs. (\ref{explicit1}) and (\ref{explicit2}) 
into Eq. (\ref{evolution}) are listed in appendix \ref{appendix}.
If we introduce a vector notation

\begin{mathletters}
\begin{eqnarray}
{\bf u}_{1} & = & 
\left(
\begin{array}{c}
\mbox{\boldmath $\lambda$} \\
\mbox{\boldmath $\alpha$}
\end{array}
\right) \; , 
\end{eqnarray}
\begin{eqnarray}
{\bf u}_{2} & = &  
\left(
\begin{array}{c}
\mbox{\boldmath $\eta^{(1)}$} \\
\mbox{\boldmath $\eta^{(2)}$}
\end{array}
\right) \; , 
\end{eqnarray}
\begin{eqnarray}
\mbox{\boldmath $\Omega$} 
& = & 
\left(
\begin{array}{cc}
\mbox{\boldmath $\omega$} & 0 \\
0 & \mbox{\boldmath $\omega$}
\end{array}
\right) \; , 
\end{eqnarray}
\end{mathletters}
where $\mbox{\boldmath $\omega$}_{ij} = \delta_{ij} \omega_{i}$,
these can be schematically written as 

\begin{mathletters}
\begin{eqnarray}
\label{schematical_eqm}
\ddot {\bf u}_{1} + \mbox{\boldmath $\Omega$}^{2} \cdot {\bf u} _{1} =  
{\bf v} \left( {\bf u}_{1}, {\bf u}_{2} \right) \; , 
\end{eqnarray}
\begin{eqnarray}
\ddot {\bf u}_{2} + \mbox{\boldmath $\Omega$}^{2} \cdot  {\bf u}_{2} =  
{\bf w} \left( {\bf u}_{1}, {\bf u}_{2} \right) \; ,
\end{eqnarray}
\end{mathletters}
with $\bf v$ and $\bf w$ nonlinear vector functions of their 
arguments.
To find the excitation frequencies of the modes, we have to linearize
these equations around the
equilibrium value $(\bar {\bf u}_{1}, \bar {\bf u}_{2})$ 
and subsequently determine the  
eigenvectors and eigenvalues of the linearized problem.
From the explicit expressions in 
appendix \ref{appendix}, it is easily
seen that the linearized equations for $\delta {\bf u}_{1}$
and $\delta {\bf u}_{2}$ decouple. 
The resulting excitation frequencies
for the in-phase and out-of-phase 
monopole, quadrupole and dipole, or Kohn modes 
are presented and discussed in section
\ref{section5} below. 
\section{LINEAR RESPONSE}
\label{section4}
In order to compare our theoretical results with experiment,
it turns out to be important to understand what is measured experimentally
when the external trapping potential is perturbed to excite the
collective modes of the Bose-condensed gas. It is possible that
a periodic modulation of the trapping frequency will excite more than one mode
when it is not exactly on resonance. The question is then which
mode is most likely to be seen experimentally. 
To answer this question we have  
to study the linearized response of the gas to 
a periodic perturbation of the trapping frequencies

\begin{eqnarray}
\omega_{i} \rightarrow \omega_{i} + 
\delta \omega_{i} e^{i \omega t} \; ,
\end{eqnarray}
leading to 

\begin{eqnarray}
V_{ext}({\bf x}) & \rightarrow & V_{ext}({\bf x}) + 
\delta V_{ext}({\bf x}) e^{i \omega t} \; ,
\end{eqnarray}
and 
\begin{eqnarray}
n({\bf x},t) & \rightarrow & 
n({\bf x}) + \delta n({\bf x}) e^{i \omega t} \; . 
\end{eqnarray}
A quantity that characterizes the response to such a perturbation is 
the time averaged work done by the perturbation \cite{jackson},

\begin{eqnarray}
\label{work}
W = \frac{1}{2} 
\int d{\bf x} \;  
\delta V_{ext}({\bf x}) \delta n^{*}({\bf x}) \; ,
\end{eqnarray}
where the asterix denotes the complex conjugate.
An explicit expression for the time-averaged work
is found  by linearizing the density profiles around equilibrium,
by putting $\lambda_{i} = \bar \lambda_{i} + 
\delta \lambda_{i} e^{i \omega t}$ and
$\alpha_{i} = \bar \alpha_{i} + \delta \alpha_{i} e^{i \omega t}$.
If we insert the resulting expression for
$\delta n = \delta n_{0} + \delta n'$
into Eq. (\ref{work}), we get
\begin{eqnarray} 
W & = & 
\sum_{i} m \omega_{i} \delta \omega_{i}
\left[
\bar \lambda_{i} \delta \lambda_{i}^{*} \langle x_{i}^{2} \rangle_{c} +
\bar \alpha_{i} \delta \alpha_{i}^{*} \langle x_{i}^{2} \rangle_{nc}
\right] \; .
\end{eqnarray}
To calculate the work done by the perturbation,
we thus need to know the response of the scaling parameters
to a perturbation of the external potential.

After linearizing the equations of motion 
Eq.~(\ref{schematical_eqm}), with ${\bf u}_{2} = 0$,
to first order in $\delta \omega_{i}$ and $\delta {\bf u}_{1}$,
the resulting equation of motion for the
fluctuation $\delta {\bf u}_{1}$, reads  

\begin{eqnarray}
\label{linearized}
-\omega^{2} \delta {\bf u}_{1} + 
\mbox{\boldmath $\Omega$}^{2} \cdot 
\delta {\bf u}_{1} & = &
\left[ \nabla_{u_{1}} {\bf v}
\right] \cdot  \delta {\bf u}_{1}
- 2 \mbox{\boldmath $\Omega$} \cdot  
\delta \mbox{\boldmath $\Omega$} 
\cdot \bar {\bf u}_{1} 
\; . 
\end{eqnarray}
The partial derivative of ${\bf v}$ with
respect to ${\bf u}_{1}$ is to be evaluated in the equilibrium point.
These linearized equations of motion are easily solved by

\begin{eqnarray}
\delta {\bf u}_{1} = \sum_{n}
\frac{a_{n}}{\omega^{2}-\omega_{n}^{2}} \; {\bf u}_{1}^{(n)} \; ,
\end{eqnarray}
where ${\bf u}_{1}^{(n)}$ denote
the normalized eigenvectors of 
the homogeneous part of 
Eq. (\ref{linearized}) and $a_{n} = 
2 {\bf u}_{1}^{(n)} \cdot
\mbox{\boldmath $\Omega$} \cdot 
\delta \mbox{\boldmath $\Omega$}\cdot 
\bar {\bf u}_{1}$.
The time averaged work done can now be expressed as 

\begin{eqnarray}
W = \sum_{n} \frac{b_{n}}{\omega^{2}-\omega_{n}^{2}} \; ,
\end{eqnarray}
where the residue $b_{n}$ is given by

\begin{eqnarray}
b_{n} & = & \sum_{i} m \omega_{i} \delta \omega_{i} a_{n}
\left[
\bar \lambda_{i} \lambda_{i}^{(n)} \langle x_{i}^{2} \rangle_{c} +
\bar \alpha_{i} \alpha_{i}^{(n)} \langle x_{i}^{2} \rangle_{nc}
\right] \; .
\end{eqnarray}

In reality, the eigenmodes are damped.
Therefore, in a theory that includes damping,
the poles $\omega_{n}$ have an imaginary part
and $W$ is always finite with a maximum at $\omega=\omega_{n}$.
Hence, a resonance would occur if the system were driven with
that frequency. A measure for the strength with which 
this mode is excited, is the residue $b_{n}$.
Also, the time-averaged work would acquire an imaginary part, 
which would determine the power absorption.
We have calculated the residue $b$ for two particular modes
as a function of temperature 
and the results are presented in the next section.
\section{RESULTS}
\label{section5}
In this section we present the results
of our calculations  
and if possible compare them with 
experimental data.
In principle our approach determines $12$ modes of the gas.
In an axially symmetric situation, 
they correspond to the in-phase and out-of-phase versions
of three Kohn modes, two monopole modes and one quadrupole mode.
These in-phase and out-of-phase modes are the collisionless
analogue of the hydrodynamic first and second sound modes 
\cite{zaremba,shenoy,kavoulakis1}.
We have calculated the excitation frequencies 
of these modes for the parameters of the 
experiments with $^{87}$Rb \cite{jin2}, $^{23}$Na \cite{stamper-kurn} 
and $^{7}$Li \cite{bradley}.
Furthermore, 
for the experiments with $^{87}$Rb, we have calculated 
the residue $b$ mentioned in the previous section
for the in-phase and out-of-phase monopole modes.

\begin{figure}
\psfig{figure=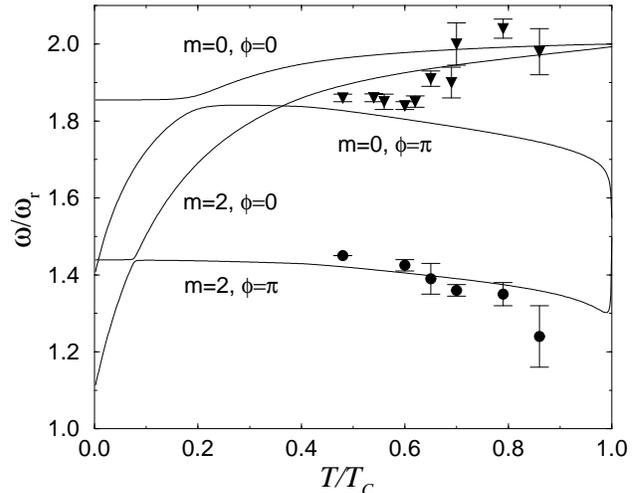}
\caption{\narrowtext
The in-phase and out-of-phase $m=0,2$ modes as a function of $T/T_{c}$
for the experimental conditions of Jin {\it et al.} [10].
The relative phase of the density profiles of the condensed 
and noncondensed atoms is denoted by $\phi$. 
Also included are the experimental results 
for the $m=0$ (triangles) and  $m=2$ (circles) modes 
found in these experiments. \label{one}}
\end{figure}

\begin{figure}
\psfig{figure=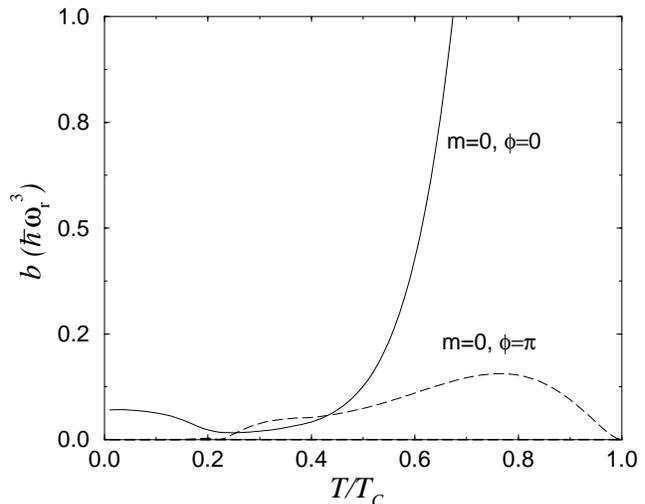}
\caption{ \narrowtext
The residue $b$ for the modes $m=0, \phi=0$ (solid line)
and $m=0, \phi=\pi$ (dotted line) as a function of $T/T_{c}$
for the experimental conditions of Jin {\it et al.} [10]. \label{two}} 
\end{figure}

From the results presented in Fig. \ref{one}, it is clear
that the temperature dependence
of the out-of-phase $m=2$ mode is in reasonable
agreement with that of the $m=2$ mode of the $^{87}$Rb experiments.
Furthermore, the experimental data for the 
$m=0$ mode goes to the correct non-interacting limit
near $T_{c}$,
where it coincides with our theoretical curve 
for the in-phase $m=2$ mode.
In addition, between $0.7$ and $0.6 \; T_{c}$ the experimental
data drops to the zero-temperature 
limit $(10/3)^{1/2} \; \omega_{r}$ \cite{stringari},
where it coincides with our theoretical curve 
for the out-of-phase $m=2$ mode.
It is important to note that in obtaining these results, we have
included the effect of evaporative cooling
by fitting the total number of particles 
to the experimental results of Jin {\it et al.}
\cite{jin2}. 
Qualitatively, it appears that the strong 
temperature dependence found experimentally
might be due to the fact that one simultaneously excites
both the in-phase and out-of-phase $m=0$ modes \cite{cornell}.
To explore this possibility we have calculated
the residue $b$ for these two modes. 
As shown in Fig. \ref{two}, there is a clear crossover at about
$0.5 \; T_{c}$, after which the value for the residue of 
the in-phase $m=0$ mode shoots up.
Therefore, the experimental data
might actually be due to the excitation of two modes, and as 
a function of temperature one crosses over
from exciting mainly one mode to exciting mainly 
the other.
Furthermore, Fig. \ref{two} shows that the 
out-of-phase $m=0$ mode 
cannot be seen experimentally below $0.2 \; T_{c}$, 
because here the mode is very difficult to excite.
Finally, we note that it might be possible to 
observe the two  additional modes present
in our calculation experimentally.
Whether this is possible depends
on the overlap of these modes with the 
applied perturbation, 
and on the damping of the modes, which we have
neglected . 

As mentioned above, our theory reproduces the correct 
excitation frequencies near the critical temperature
and near zero temperature. In principle, sufficiently far
below the critical temperature there is a `dimple' 
in the noncondensate density profile due to the presence of a condensate.
This `dimple' is not taken into account in our
dynamical scaling {\it ansatz}. 
However, it is incorrect to include this `dimple' 
by simply modifying the scaling profile,
because we are using a one parameter scaling {\it ansatz},
Consider for example an out-of-phase
mode of the condensate and the noncondensate.

Because the `dimple' is caused by the mean-field 
interaction with the condensate, it also has to move 
out-of-phase with the exterior part of the noncondensed cloud. This implies
that there are two length scales that determine the dynamics of the
noncondensate. Hence, modifying the scaling profile $\Phi_{F}$, for example
by taking the exact equilibrium solution for the noncondensate 
density, can actually give worse results. 
Moreover, neglecting
the `dimple' in the density profile of the 
noncondensed cloud is certainly correct
near the critical temperature. 
A qualitative reason why the theory still seems
to agree well with experiment, is 
that a `dimple' in the noncondensate density 
on the one hand increases the effective mass of the condensate,
which reduces the excitation frequency,
but on the other hand lowers the mean-field interaction,
which increases the excitation frequency. Apparently, these
two effects almost cancel each other.  

\begin{figure}[h]
\psfig{figure=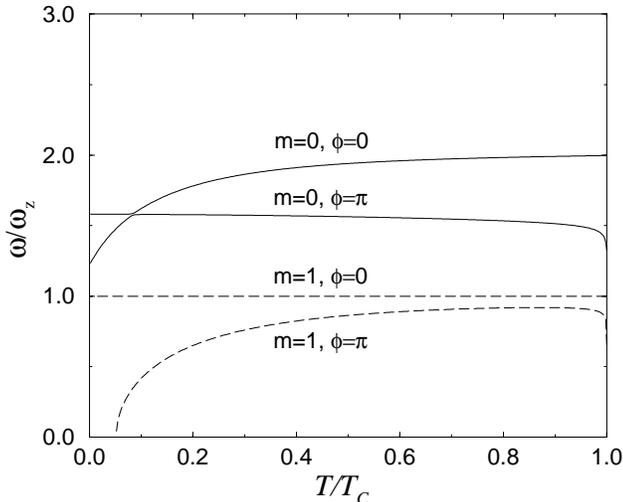}
\caption{
\label{four}
\narrowtext
The in-phase and out-of-phase $m=0$ (solid lines), and Kohn 
(dashed lines) modes as a function of $T/T_{c}$
for the experimental conditions
of Stamper-Kurn {\it et al.} [9],
with a total number of particles $N = 40 \times 10^{6}$}
\end{figure}

In Fig. \ref{four}, 
we show the calculated 
in-phase and out-of-phase 
monopole and Kohn modes
for the experimental parameters
of Stamper-Kurn {\it et al.} \cite{stamper-kurn}. 
The Kohn modes
are exactly present in our theory, which
can be seen by rewriting Eq. (\ref{evolution})
in terms of $\eta_{i}^{(1)} + \eta_{i}^{(2)}$ 
and $\eta_{i}^{(1)} - \eta_{i}^{(2)}$.
The out-of-phase Kohn modes have the qualitatively 
correct feature found in the experiments, that the 
frequency of the mode is shifted downward
with respect to the trapping frequency.
Experimentally the shift is about $5 \%$. Instead, we
find a reduction of about $10 \%$.
The fact that the out-of-phase dipole mode becomes unstable
at $T/T_{c} \approx 0.05$ 
can be understood by realizing that our equilibrium profile
is always centered around the origin. Because we do not include
the `dimple' in the noncondensate density profile this implies
that at a certain temperature, it is energetically favorable for
the noncondensed cloud to shift it's center outward. 
Because the instability occurs only at very low temperatures,
this artifact of our {\it ansatz} is not of importance for our purposes.

\begin{figure}[h]
\psfig{figure=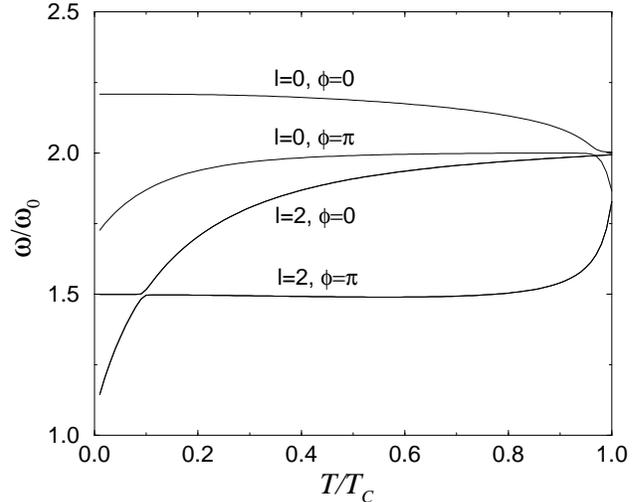}
\caption{
\label{five}
\narrowtext
The in-phase and out-of-phase $m=0,2$ 
modes as a function of $T/T_{c}$
for an isotropic trap $\omega_{x}=\omega_{y}=\omega_{z}=\omega_{0}=200 Hz$
and with a total number of atoms $N = 2000$.}
\end{figure}

To be able to compare
our data with results found 
in the Popov calculations of Zaremba {\it et.al.} \cite{hutchinson1},
we have also looked at the case of an isotropic trap.
Interestingly, our results, which are shown in Fig. \ref{five},
are quite similar to theirs.
Moreover, also when analyzing the temperature dependence
of the mode frequencies by means 
of a temperature dependent effective interaction,
the results found in the isotropic case are very
similar to these results \cite{hutchprivate}.
This suggests that the interesting temperature dependencies
are related to the anisotropy of the
external trapping potential. 

\begin{figure}[h]
\psfig{figure=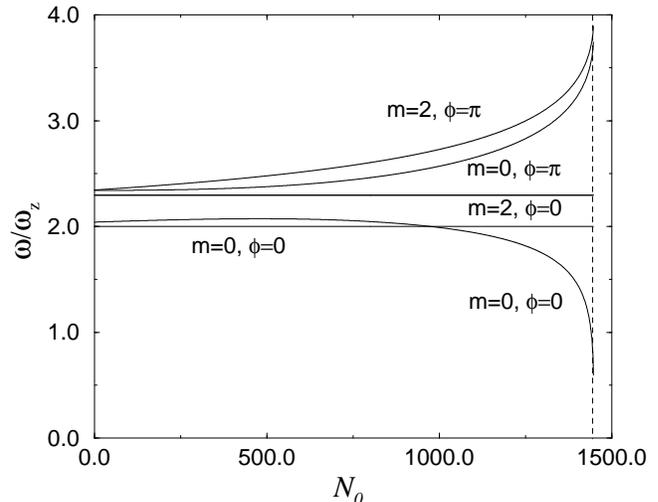}
\caption{
\label{six}
\narrowtext
The in-phase and out-of-phase $l=0,2$ modes (solid lines) and Kohn 
(dashed lines) modes at constant temperature $T=300 nK$, 
as a function of the number of
atoms in the condensate, 
for the experimental conditions of Bradley {\it et al.} [2].
The condensate collapses
for $N_{0}=1447$.}
\end{figure}

Finally, as shown in Fig. \ref{six},
we have also calculated the mode frequencies for 
the experimental conditions of Bradley {\it et al.} \cite{bradley}
that apply to the case of a negative scattering length.
It should be noted that our approach is 
particularly suited for a discussion of this case,
because the mean-field interaction of the condensate 
is at most comparable to the energy splitting in the trap,
due to the intrinsic instability of the condensate to 
collapse \cite{stoof97}. As expected,
the maximum number of condensate particles is slightly shifted
downwards by the mean-field interaction of the noncondensate
with the condensate. 
At a temperature of $300 \; nK$, we find a shift of about $3 \%$,
whereas more accurate calculations find a shift
of about $5 \%$ \cite{houbiers}.
In addition, the dynamical treatment
of the noncondensed cloud seems to have little effect on the
mode frequencies. This indicates that it is a good approximation
to treat the collapse 
purely in terms of the condensate even at nonzero temperatures \cite{sackett}.
\section{CONCLUSION AND OUTLOOK}
\label{section6}
If we interpret the 
temperature dependence of the $m=0$ mode found in the
JILA experiments as due to the excitation of two modes
instead of one,
there seems to be reasonable agreement between our theoretical 
predictions and the experimental results. 
We should mention, however, that
our results are sensitive to the explicit form of the {\it ansatz}. 
Indeed, work done by the authors in collaboration with 
E. Zaremba, shows that using for the {\it ansatz} a 
numerical solution for the
condensate and noncondensate equilibium density profiles
in the Popov approximation,
leads to qualitatively different results. 
However, as mentioned above, 
we believe this to be due to the one-parameter
nature of the {\it ansatz} used. It is a matter of future
investigation to resolve this problem, possibly by including more
variational parameters into the density profile.
Alternatively, we can perform a RPA-calculation of the mode 
frequencies \cite{minguzzi}, which is essentially equivalent
to finding the modes of the collisionless Boltzmann equation
coupled to the nonlinear Schr\"odinger equation 
exactly \cite{kadanoff}.  
Finally, we can in principle also calculate
the damping of these modes within our approach, by including
the collision terms in the Boltzmann equation \cite{kavoulakis2}. 
Work to implement these ideas is in progress.
\section*{ACKNOWLEDGMENTS}
It's a pleasure to thank E. Zaremba for many 
fruitful discussions and for collaborating in the 
calculation of the mode frequencies 
with equilibrium density profiles as our {\it ansatz}.
Furthermore, we acknowledge
illuminating comments by E.A. Cornell, F. Langeveld and C.J. Pethick.
\appendix
\section{EQUATIONS OF MOTION}
\label{appendix}
We list here the equations of motion
for the variational parameters 
$\{ \lambda_{i} \}$, $\{ \alpha_{i} \}$, $\{ \eta_{i}^{(1)} \}$ and 
$\{ \eta_{i}^{(2)} \}$
resulting from inserting
the explicit form of
the condensate wave function and the noncondensate 
distribution function Eqs. (\ref{explicit1}) and (\ref{explicit2}) 
into Eqs. (\ref{evolution}).  
The average harmonic oscillator length is
defined by $\bar l=\sqrt{\hbar  /m \bar \omega}$, where
$\bar \omega = (\omega_{1} \omega_{2} \omega_{3})^{1/3}$.
The thermal wavelength is given by $\Lambda_{th} =
\sqrt{2 \pi \hbar^{2} / m k_{B} T}$ and $\zeta(4) \approx 1.082$.
Note that in equilibrium $\eta_{i}^{(1)}=\eta_{i}^{(2)}=0$
and that the equations for 
$\{ \lambda_{i}, \alpha_{i} \}$ and 
$\{ \eta_{i}^{(1)}, \eta_{i}^{(2)} \}$
decouple when they are linearized around this equilibrium.
\end{multicols}
\widetext
\begin{eqnarray}
\label{evolution2}  
\frac{\ddot \eta_{i}^{(1)}}{\omega_{i}^{2}} + \eta_{i}^{(1)} & = &
\frac{4 \zeta(3)^{-1} }{\pi^{3} \sqrt{2}}
\left( \frac{N' a}{\bar{l}} \right)
\left(
\frac{\Lambda_{th}}{\bar{l}}
\right)^{5} 
\sum_{n=1}^{\infty} \frac{1}{n^{1/2}} 
\nonumber \\
& & 
\prod_{j} \left[ \frac{1}{n \beta \hbar \omega_{j} \lambda_{j}^{2} + 
2 \alpha_{j}^{2} \bar \alpha_{j}^{2}} \right]^{\frac{1}{2}}
\frac{\eta^{(1)}_{i}-\eta^{(2)}_{i}}
{n \beta \hbar \omega_{i} \lambda_{i}^{2} + 
2 \alpha_{i}^{2} \bar \alpha_{i}^{2}} 
\exp \left( \sum_{i} 
\frac{-n \beta m \omega_{i}^{2} \left[
\eta^{(1)}_{i}-\eta^{(2)}_{i} \right]^{2}}
{n \beta \hbar \omega_{i} \lambda_{i}^{2} + 
2 \alpha_{i}^{2} \bar \alpha_{i}^{2}} \right) \; .
\\ 
\frac{\ddot \eta_{i}^{(2)}}{\omega_{i}^{2}} + \eta_{i}^{(2)} & = &
\frac{4 \zeta(3)^{-1} }{\pi^{3} \sqrt{2}}
\left( \frac{N_{0} a}{\bar{l}} \right)
\left(
\frac{\Lambda_{th}}{\bar{l}}
\right)^{5} 
\sum_{n=1}^{\infty} \frac{1}{n^{1/2}} 
\nonumber \\
& & 
\prod_{j} \left[ \frac{1}{n \beta \hbar \omega_{j} \lambda_{j}^{2} + 
2 \alpha_{j}^{2} \bar \alpha_{j}^{2}} \right]^{\frac{1}{2}}
\frac{\eta^{(2)}_{i}-\eta^{(1)}_{i}}
{n \beta \hbar \omega_{i} \lambda_{i}^{2} + 
2 \alpha_{i}^{2} \bar \alpha_{i}^{2}} 
\exp \left( \sum_{i} 
\frac{
-n \beta m \omega_{i}^{2} \left[
\eta^{(2)}_{i}-\eta^{(1)}_{i} \right]^{2}}
{n \beta \hbar \omega_{i} \lambda_{i}^{2} + 
2 \alpha_{i}^{2} \bar \alpha_{i}^{2}} \right) \; .
\\ 
\frac{\ddot \lambda_{i}}{\omega_{i}^{2}} + \lambda_{i} & = &
\frac{1}{\lambda_{i}^{3}} + \sqrt{\frac{2}{\pi}}
\frac{N_{0} a}{\bar{l}}
\left(
\frac{l_{i}}{\bar{l}}
\right)^{2} 
\prod_{j} \left[ \frac{1}{\lambda_{j}} \right]
\frac{1}{\lambda_{i}} + 
\nonumber \\
& & 
\frac{4 \zeta(3)^{-1} }{\pi^{3} \sqrt{2}}
\left( \frac{N' a}{\bar{l}} \right)
\left(
\frac{\Lambda_{th}}{\bar{l}}
\right)^{5} 
\sum_{n=1}^{\infty} \frac{1}{n^{1/2}} 
\prod_{j} \left[ \frac{1}{n \beta \hbar \omega_{j} \lambda_{j}^{2} + 
2 \alpha_{j}^{2} \bar \alpha_{j}^{2}} \right]^{\frac{1}{2}}
\frac{\lambda_{i}}
{n \beta \hbar \omega_{i} \lambda_{i}^{2} + 
2 \alpha_{i}^{2} \bar \alpha_{i}^{2}} 
\nonumber \\
& & 
\left[ 1 +
\frac{2 n \beta m \omega_{i}^{2} \left[
\eta^{(1)}_{i}-\eta^{(2)}_{i} \right]^{2}}
{n \beta \hbar \omega_{i} \lambda_{i}^{2} + 
\alpha_{i}^{2} \bar \alpha_{i}^{2}} \right] 
\exp \left( \sum_{i} 
\frac{
-n \beta m \omega_{i}^{2} \left[
\eta^{(1)}_{i}-\eta^{(2)}_{i} \right]^{2}}
{n \beta \hbar \omega_{i} \lambda_{i}^{2} + 
2 \alpha_{i}^{2} \bar \alpha_{i}^{2}} \right) \; .
\\
\frac{\ddot \alpha_{i}}{\omega_{i}^{2}} + \alpha_{i} & = &
\frac{1}{\alpha_{i}^{3}} + 
\frac{1}{2 \pi^{3}}
\left(
\frac{\Lambda_{th}}{\bar{l}}
\right)^{5} 
\left( \frac{N 'a}{\bar{l}} \right)
\frac{
\left[ \sum_{n,m=1}^{\infty} n^{-1/2} m^{-3/2} (n+m)^{-5/2} \right]}{
\zeta(3) 
\zeta(4)}
\left[ 
\prod_{j} \frac{1}{\alpha_{j} \bar \alpha_{j}} \right]
\frac{1}{\alpha_{i} \bar \alpha_{j}^{2}} + 
\nonumber \\
& &
\frac{4 \zeta(4)^{-1} }{\pi^{3} \sqrt{2}}
\left( \frac{N_{0} a}{\bar{l}} \right)
\left(
\frac{\Lambda_{th}}{\bar{l}}
\right)^{5} 
\sum_{n=1}^{\infty} \frac{1}{n^{3/2}} 
\prod_{j} \left[ \frac{1}{n \beta \hbar \omega_{j} \lambda_{j}^{2} + 
2 \alpha_{j}^{2} \bar \alpha_{j}^{2}} \right]^{\frac{1}{2}}
\frac{\alpha_{i}}{n \beta \hbar \omega_{i} \lambda_{i}^{2} + 
2 \alpha_{i}^{2} \bar \alpha_{i}^{2}} 
\nonumber \\
& & 
\left[ 1 +
\frac{2 n \beta m \omega_{i}^{2} \left[
\eta^{(2)}_{i}-\eta^{(1)}_{i} \right]^{2}}
{n \beta \hbar \omega_{i} \lambda_{i}^{2} + 
2 \alpha_{i}^{2} \bar \alpha_{i}^{2}} \right] 
\exp \left( \sum_{i} 
\frac{
-n \beta m \omega_{i}^{2} \left[
\eta^{(2)}_{i}-\eta^{(1)}_{i} \right]^{2}}
{n \beta \hbar \omega_{i} \lambda_{i}^{2} + 
2 \alpha_{i}^{2} \bar \alpha_{i}^{2}} \right) \; .
\end{eqnarray}
\begin{multicols}{2}

\end{multicols}
\end{document}